\documentclass[preprint]{jpsj2}
\usepackage{epsf}
\usepackage{graphicx}

\newcommand{\citePRB}[3]{Phys.\ Rev.\ B {\bf #1} (#2) #3}
\newcommand{\citePRL}[3]{Phys.\ Rev.\ Lett. {\bf #1} (#2) #3}
\newcommand{\citeRMP}[3]{Rev.\ Mod.\ Phys. {\bf #1} (#2) #3}
\newcommand{\citeJPSJ}[3]{J.\ Phys.\ Soc.\ Jpn. {\bf #1} (#2) #3}

\newcommand{\citeEPJB}[3]{Eur.\ Phys.\ J. B {\bf #1} (#2) #3}

\newcommand{\del}{\partial}

\def\runauthor{Masatoshi {\sc Imada}
}

\title{Quantum Mott Transition and Multi-Furcating Criticality}

\author{Masatoshi Imada$^{1,2}$}
\inst
{{$^1$}
Institute for Solid State Physics, 
University of Tokyo, 5-1-5, Kashiwanoha, Kashiwa, Chiba 277-8581\\
{{$^2$}PRESTO, Japan Science and Technology Agency}
}

\abst{
Phenomenological theory of the Mott transition is presented.  When the critical temperature of the Mott transition is much higher than the quantum degeneracy temperature, the transition is essentially described by the Ising universality class. Below the critical temperature, phase separation or first-order transition occurs.  However, if the critical point is involved in the Fermi degeneracy region, a marginal quantum critical point appears at zero temperature.  The originally single Mott critical point generates subsequent many unstable fixed points through various Fermi surface instabilities induced by the Mott criticality characterized by the diverging charge susceptibility or doublon susceptibility. This occurs in marginal quantum-critical region.  Charge, magnetic and superconducting instabilitites compete severely under these critical charge fluctuations.  The quantum Mott transition triggers multi-furcating criticality, which goes beyond the conventional concept of multicriticality in quantum phase transitions.  
Near the quantum Mott transition, the criticality generically drives growth of inhomogeneous structure in the momentum space with singular points of flat dispersion on the Fermi surface.  The singular points determine the quantum dynamics of the Mott transition by the dynamical exponent $z=4$.  We argue that many of filling-control Mott transitions are classified to this category.
Recent numerical results as well as experimental results on strongly correlated systems including transition metal oxides, organic materials and $^3$He layer adsorbed on a substrate are consistently analyzed especially in two-dimensional systems.  The mechanism of cuprate high-$T_c$ superconductivity is also discussed in the light of the present insight and interpreted from the multi-furcation instability.   
}
\kword{Ginzburg-Landau theory, quantum phase transition, Mott transition, multi-furcation, two-dimensional Hubbard model, high-Tc superconductivity, quantum critical phenomena, marginal critical point, high-Tc superconductivity, phase separation}

\begin{document}
\sloppy
\maketitle

\date{January 15, 2004}
%%%%%%%%%%%%%%%%%%%%%%%%%%%%%%%%%%%%%%%%%%%%
%% MAINMATTER
%%%%%%%%%%%%%%%%%%%%%%%%%%%%%%%%%%%%%%%%%%%%

\section{Introduction}
% ‹à'®â‰'Ì"]ˆÚ'Ɖt'Š‹C'Š"]ˆÚA'Ώ̐«'Ì"j'ꂪ'È'¢'Æ'«Bƒoƒ"ƒhâ‰'Ì"]ˆÚ'à'Ώ̐«'Ì"j'ê–³'¢'ªAA
Recently, enormous number of experimental results have been accumulated on properties of metals near correlated insulators such as the Mott insulator~\cite{RMP}.  In these results, the nature shows diverse properties depending on subtle differences in systems and ranging from strongly renormalized Fermi liquid, charge order, magnetic order to superconductivity at low temperatures.  More complicated and entangled phase diagrams and very sensitive responses to external perturbations are also observed.  In some cases, severe competitions of multiple orders are found.  Inhomogeneous structure at the Fermi surface is also widely observed particularly in the high-$T_c$ cuprates.  This includes an emergence of flat dispersion in special region of the Brillouin zone as in the $(\pi,0)$ and $(0,\pi)$ regions in the high-$T_c$ cuprates~\cite{RMP,Shen,Ino00}.  Spatially inhomogeneous structure is also widely recognized in the cuprates~\cite{Pan} as well as in manganaites~\cite{Mn1,Mn2}.  Such dramatic competitions and self-organized structure formations in real and momentum spaces have not been expected in weakly correlated electron systems.   

Intensive studies on theoretical models of strongly correlated electrons have also revealed subtlety of the results on stabilities of phases and controversies depending on approaches, approximations and the models themselves~\cite{RMP}.  These controversies and diversities both in theories and experiments consistently show not necessarily the failure of the present theoretical treatments for each problem but the emergence of an underlying new class of phenomena and necessity for a new concept in physical systems in a region of strong electron correlation to reproduce the diversity and sensitivity.    

%'±'̘_•¶'Ì–Ú"I
In this paper, a phenomenological theory of the Mott physics emerging near the Mott insulator is studied and insights from recent experimental results together with numerical results of theoretical models are analyzed.  We show that the Mott transition of Fermion systems and its criticality provides a new type of quantum phase transitions, where a classical Ising-type transition is transformed to a quantum one in a unique way.  This unique feature naturally accounts for diversity of phenomena with underlying strong competition of orders, emergence of structure in real and momentum spaces. 

When the critical temperature of the Mott transition is much higher than the energy scale of the quantum degeneracy, the transition at the critical point retains a classical nature, which is analogous to the gas-liquid transition.
Around the critical point, the diverging and critical density fluctuation is expected.
At low temperatures, the transition is of the first order and strong quantum fluctuations do not appear.  

% —ÊŽqŒø‰Ê'Ƃ̂'Ȃª'è
The gas-liquid type transition (or phase separation) in quantum systems was studied by Blume, Emery and Griffiths~\cite{BEG} in the context to understand the phase diagram of  $^3$He-$^4$He mixture with the superfluid transition of $^4$He being involved.  For Fermion systems, Castellani {\it et al.}~\cite{Castellani} extended this study to the Fermion Hubbard model by including magnetic degrees of freedom and the doubly occupied sites as well.  The metal-to-insulator transition may be treated by the analogy to the gas-liquid transition in this framework and Kotliar {\it et al.}~\cite{Kotliar,Kotliar2} indeed derived the Ising-type transition within the framework of the dynamical mean field theory~\cite{Metzner,Georges}.  The critical temperature was high enough so that it was recognized to be equivalent to the classical Ising-type transition and quantum effects are rather irrelevant.  
A fundamental open problem still remains in the issue how quantum effects alter the criticality of the conventional gas-liquid or bainary-alloy transitions of the Ising type.    

In Fermion systems, if the critical temperature at the termination point of the first-order transition is suppressed, the diverging density fluctuations inherent at the critical point of the gas-liquid transition become involved in the quantum Fermi degeneracy region.  The Fermi degeneracy by itself generates various instabilities called the Fermi surface effects.  Then the coexistence of these two becomes an intriguing issue.

The purpose of this paper is to show that the coexistence of the Fermi degeneracy and the critical density fluctuations yields a new type of quantum criticality. The Mott transition offers a salient example of this issue.
At this quantum criticality, the mother transition, namely the Mott transition triggers various daughter instabilities and quantum criticalities such as charge, magnetic, and superconducting transitions. In other words, at the unstable fixed point of the root transition, namely the Mott transition, because of its criticality itself, it generates many other unstable fixed points simultaneously if it coexists with a quantum degeneracy.    The coexistence also induces inhomogeneity at the Fermi surface in the momentum space as well as in real  space.  
%The resultant competitions of orders and its remaining fluctuations introduce a large residual entropy even at low temperatures.  
As far as the author knows, this type of {\it multi-furcating} or cascading quantum criticality originating from one quantum phase transition has never been considered before and offers a new type of physics beyond the conventional theory of multicritical phenomena in quantum phase transitions as we discuss later.  The critical region may not be characterized by simple critical exponents but by possible cascading and hierarchical structure formations.  The phenomena must first be understood as a whole; not by individual competitions and transitions but from a synergetic viewpoint. 
Quantum critical phenomena have been intensively studied in case of the magnetic critical point, where the vanishing magnetic critical temperature in metals causes non-Fermi liquid behavior and large fluctuations in the critical region~\cite{Moriya,Hertz,RMP}.  
In the present case, quantum criticality of the Mott transition generates a quite different type through diverging charge fluctuations, where the critical temperature as the end point of a first-order transition of the electronic (or excitonic) density is suppressed and a marginal critical point appears at zero temperature.  

The applicability of this new class of quantum criticality is not confined to the Mott transition but may also be extended to other type of first-order transitions such as the charge order transitions when their critical points are involved in the Fermi degeneracy region.  Such examples may also be found near the first-order chage of the valence in heavy fermioin compounds.
Valence fluctuations in heavy fermion compounds and density fluctuations for charge orders in organic and transition metal compounds offer such examples.

In Sec. 2, we summarize the general scheme of the Ginzburg-Landau (GL) expansion within the classical framework.  In Sec. 3, two-parameter expansion for the Mott transition is formulated.  In Sec. 4, we discuss comparison with results of recent numerical and microscopic calculations as well as comparison with experimental results.  We consider quantum effects in Sec. 5.  A two-fluid model of the Mott transition is presented in Sec. 6.  Section 7 is devoted to a consequence of the coexistence of the Fermi degeneracy and the Mott criticality, where multi-furcation instability is considered. In Sec.8, insights into experimental indications are discussed.  A mechanism of high-temperature superconductivity is also considered.

% 1ŽŸ"]ˆÚA@ŒÃ"T—̈æA—ÕŠE"_‹ß–T
% "d‰×Š´Žó—¦Aƒ_ƒuƒƒ"Š´Žó—¦,"d‰×'ä'炬'Ì"­ŽU''å
\section{Ginzburg-Landau (GL) Expansion}
\subsection{Classical GL Picture}
At finite temperatures, a metal and an insulator may be separated in the phase diagram if a first-order transition takes place. The first-order transition  should terminate at the critical end point at a finite temperature $T_c$.  To understand the origin of the first-order transition, the equivalence with the   gas-liquid transitions or the binary-alloy phase separation has been identified~\cite{Kotliar,Kotliar2}.   

In the mean-field Ginzburg-Landau (GL) expansion of the conventional gas-liquid transition, the free energy is expressed by coarse-grained variables of the natural order parameter, namely, the density $n$ as 
\begin{equation}
F = \frac{1}{2}a(T-T_c)n^2+\frac{1}{4}bn^4
\label{1} 
\end{equation} 
with $a$ and $b$ being constants.
Here the density is measured from the critical density.
At the critical temperature $T=T_c$, in this mean-field theory, the charge susceptibility (compressibility) $\chi_c \equiv (d^2F/dn^2)^{-1}$ diverges as $\chi_c = 1/(3bn^2)$ at $n=0$, which results in diverging and critical density fluctuations.
As a function of temperature, the charge susceptibility also diverges at the critical density
as $\chi_c \propto (T-T_c)^{-\gamma}$ with the mean-field exponent $\gamma=1$.
Below $T_c$, the first-order transition between gas and liquid occurs between $n=\sqrt{a/b}$ and $n=-\sqrt{a/b}$.

% ƒ'ƒbƒg"]ˆÚ'Ì'f'k—˜_ADMFT, ƒCƒWƒ"ƒO••Õ«'̃Œƒrƒ…[
%\subsection{GL Theory for the Mott Transition}
The Mott transition has been considered in the GL expansion in the dynamical mean-field theory~\cite{Kotliar,Kotliar2}. Essentially, the dynamical mean-field theory reproduces the Ising-type transition.  The natural order parameter is identified as the single-particle local Green's function itself, which represents the transition between localized and itinerant nature of electrons at zero frequency.  The transition is well defined at finite temperatures when the structure of the low-frequency Green's function shows a jump as the first-order transition. 

%ƒnƒo[ƒh–ÍŒ^'Ì"±"üA'Q''̃‹[ƒg'̃'ƒbƒg"]ˆÚ'Ɛ§Œäƒpƒ‰ƒƒ^
\subsection{Two Routes of Mott Transition}
Microscopically, the driving force of the Mott transition arises from the competition of the kinetic energy and the interaction energy of electrons.  The $N$-site Hubbard model defined by 
\begin{eqnarray}
{\cal H} & = &{\cal H}_t + \sum_i H_{Ui} -\mu M N \label{3.1} \\
{\cal H}_t & = & -\sum_{\langle
ij\rangle}t_{ij}(c^{\dagger}_{i\sigma}c_{j\sigma}+h.c.) \label{3.2} \\
\end{eqnarray} 
and
\begin{equation}
{\cal H}_{Ui} = U (n_{i\uparrow}-\frac{1}{2})(n_{i\downarrow}-\frac{1}{2})\label{3.4},
\end{equation}  
represents this competition in a simplest way.  
Here, 
$M  \equiv  \sum_{i\sigma} n_{i\sigma}/N$ and $n_{i\sigma}=c^{\dagger}_{i\sigma}c_{i\sigma}$ with the creation (annihilation) operator $c^{\dagger}_{i\sigma} (c_{i\sigma})$ of an electron at the site $i$ with the spin $\sigma$.  
The chemical potential is $\mu$ and $U$ is the onsite Coulomb repulsion. 
The competition of the kinetic and interaction energies can be controlled by 
two independent parameters, one, the bandwidth relative to the interaction and the other, the chemical potential $\mu$. Then the first route of the Mott transition is realized by changing the ratio $t_{ij}/U$. The first route, the bandwidth-control transition may also be driven by changing $U$ itself.  The other route is actually realized in many cases by changing the filling, which is the conjugate quantity to the control parameter $\mu$.   The natural order parameter of the transition is given by the quantity conjugate to the control parameter. In the Hubbard model Hamiltonian, the conjugate quantity to the control parameter $U$ is nothing but the avearged double occupancy defined by the ``doublon density", $n_d \equiv \langle n_{i\uparrow}n_{i\downarrow} \rangle$ as we see from Eq.(\ref{3.4}).  Therefore, $n_d$ or the averaged empty site defined by the ``holon density",  $n_h\equiv \langle (1-n_{i\uparrow})(1-n_{i\downarrow}) \rangle$ may be taken as a natural order parameter of phenomenological theory for the bandwidth-control transition at half filling. At half filling, we have the constraint $n_d=n_h$.  The natural order parameter of the second route is given by the filling $n \equiv \langle M \rangle$ or the doping concentration $\delta=1-n$, which is conjugate to the chemical potential $\mu$.
Then the GL expansion may be taken by the double expansion of two independent parameters, $\nu_s=n_d+n_h$ and $\delta_s=n_h-n_d$.

%@2ƒpƒ‰ƒƒ^'É'æ'郉ƒ"ƒ_ƒE"WŠJ
\section{Two-Parameter GL Expansion for Mott Transition}
\subsection{Framework}
% ˆø—Í'ª‹­'­'āA—ÕŠE"_'ªŒÃ"T—̈æ'É' 'é'Æ'«@Tc^d'ÆTc^n'Í"¯'¶'Å'È'¢'Æ'µ'Ä"WŠJ
The GL expansion of the free energy near the critical point may be phenomenologically constructed from the expansion in terms of two natural order parameters $\nu=\nu_s-\nu_0$ and $\delta=\delta_s-\delta_c$ measured from the critical points, $\nu=\nu_0$ and $\delta=\delta_c$. Here, the control parameter is shifted from $\nu_s$ to $\nu$ and measured from the critical value $\nu_0$.
 Near the critical point of the bandwidth-control transition, the free energy is expanded as
\begin{eqnarray}
F & = & \frac{1}{2}A_{h0}(T-T_{ch0})\delta^2 +\frac{1}{2}A_d(T-T_{cd})\nu^2 \nonumber \\ 
 & + &\frac{1}{4}B_{h0}\delta^4+\frac{1}{2}B_{hd}\delta^2\nu^2 + \frac{1}{4}B_d\nu^4 ,
\label{GL1}
\end{eqnarray}
where $A_{h0}, A_d, B_{h0}, B_d$, $B_{hd}, T_{ch0}$ and $T_{cd}$ are positive constants.  Odd order terms do not appear because of the symmetry of the critical point.
In this coarse grained form, only $\nu$ and $\delta$ are retained as variables and this form should be regarded as the result after tracing out other degrees of freedom including spins.
Since the two control parameters are independent, the critical temperatures $T_{ch0}$ and $T_{cd}$ are not necessarily the same in this formulation.
From the minimization with respect to $\nu$, 
the double occupancy at the free energy minimum is given by 
\begin{equation}
\nu=\nu_m \equiv \pm\sqrt{\frac{A_d(T_{cd}-T)-B_{hd}\delta^2}{B_d}}.
\label{GL2}
\end{equation}
for $T<T_{cd}$ and $|\delta| \leq \delta_w \equiv \sqrt{A_d(T_{cd}-T)/B_{hd}}$.
For $|\delta|>\delta_w$, $F$ is minimized at $\nu=\nu_m=0$.
One consequence of this form of free energy is that the first-order transition occurs even from metal to metal when the bandwidth is controlled in the region $|\delta| < \delta_w$ below $T_{cd}$, if this density region can be realized.
    
By taking $\nu=\nu_m$ and eliminating $\nu$, the free energy is expressed by the single parameter $\delta$ as 
\begin{equation}
F=\frac{1}{2}A_{h}(T-T_{ch})\delta^2 +\frac{1}{4}B_h\delta^4 + C,
\label{GL3}
\end{equation}
where 
\begin{equation}
A_h=A_{h0}-\frac{A_dB_{hd}}{B_d},
\label{GL4}
\end{equation}
\begin{equation}
T_{ch}=\frac{1}{A_h}(A_{h0}T_{ch0}-\frac{A_dB_{hd}T_{cd}}{B_d})
\label{GL5}
\end{equation}
and
\begin{equation}
B_{h}=B_{h0}-\frac{B_{hd}^2}{B_d}
\label{GL6}
\end{equation}
for $|\delta| \leq \delta_w$, while $\nu_m=0$ leads to $A_h=A_{h0}$, $T_{ch}=T_{ch0}$ and $B_h=B_{h0}$
for $|\delta|>\delta_w$. Here $C$ is a constant. This free energy form requires $B_h$ to be a positive constant.
This mean-field expansion is justified only near $T=T_{cd}$ and $\nu=\nu_s$, while the degeneracy structure is qualitatively correct along the first-order line below $T_{cd}$.  

%ƒoƒ"ƒh•§Œä'Å'Ì's'ƒ'Æ"­ŽU
\subsection{Bandwidth-Control Transition}
Below $T_{cd}$ at $\delta=0$, the first-order transition occurs through the jump of $\nu$ from $-\nu_m$ to $\nu_m$.  At $T_{cd}$, the ``doublon susceptibility" defined by $\chi_d\equiv[\del^2 F/\del \nu^2]^{-1}$ diverges, which means that electron-hole or excitonic density fluctuation diverges.
The mean-field exponent is given by $\chi_d=(T-T_{cd})^{-\gamma_d}$ with $\gamma_d=1$ and $\chi_d \propto \nu^{-\zeta_d}$ with $\zeta_d=2$.  The correct exponents within the classical Ising universality is given by $\gamma_d=7/4$ and $\zeta_d=14$ in two dimensions and $\gamma_d\sim 1.24$ and $\zeta_d=3.8$ in three dimensions~\cite{Goldenfeld}.
At $T=0$, the first-order bandwidth-control transition at $U=U_c$ is connected to the first-order line by putting $T=0$ in Eq.(\ref{GL1}).
Namely, the jump of $\nu$ at $U=U_c$ and $T=0$ is $\Delta \nu = 2\sqrt{A_dT_{cd}/B_d}$, although the quantitative aspect of the GL expansion becomes questionable if $\Delta \nu$ is not small.

%ƒtƒBƒŠƒ"ƒO§Œä'Å'Ì's'ƒ'Æ"d‰×Š´Žó—¦"­ŽU
\subsection{Filling-Control Transition}
When the filling is controlled, the charge susceptibility defined by $\chi_c \equiv dn/d\mu=[\del^2 F/\del \delta^2]^{-1}$ diverges at the critical point $T=T_{ch}$.  

The filling-control transition occurs at $U>U_c$, where the free energy can be expanded around the minimum $-\nu_m$ representing the Mott insulator for $\nu$.  Since $\nu$ does not have criticality away from $U_c$, the free energy can be expanded solely by $\delta$ after eliminating $\nu$.
Then the free energy has essentially the same form as eq.(\ref{GL3}).
The critical exponents are the same as the case of the bandwidth control. Namely, we obtain the mean-field exponent $\chi_c=(T-T_{ch})^{-\gamma_c}$ with $\gamma_c=1$ and $\chi_c \propto \nu^{-\zeta_c}$ with $\zeta_c=2$, while the correct exponents within the classical Ising universality is given by $\gamma_c=7/4$ and $\zeta_c=14$ in two dimensions and $\gamma_c\sim 1.24$ and $\zeta_c=3.8$ in three dimensions.

%'Š•ª—£AƒXƒsƒm[ƒ_ƒ‹A
Below $T=T_{ch}$, the phase separation generates a miscibility gap in the density, where the regions of the spinodal decomposition and nucleation are contained. The phase separation occurs because the total particle density is conserved.   However, the phase separation has a subtlety because the electrons have charge.  The real macroscopic phase separation should be suppressed because of the long-ranged part of the Coulomb interaction, which is beyond the scope of the Hubbard model. In both the nucleation and spinodal regions, the real phase separation is suppressed and may be frozen at a nanoscale regime of the phase separation as if it is frozen at a nonequilibrium state in neutral systems.  A related attempt has been made for micro phase separation of weakly charged polymer~\cite{Onuki,Spinodal}.  Nucleation and spinodal regions may show distinct frozen patterns.  In the spinodal region, typical entangled stripped pattern at an early stage of the spinodal decomposition is expected to freeze while droplet-like pattern may be seen in the nucleation region.  In the nucleation region, the phase separation would be completely suppressed when the critical nuclei radius is larger than the radius allowed from the electrostatic condition.  
However, if the inhomogeneity in the momentum space occurs as we discuss later, the suppression by the long-ranged Coulomb force may be relaxed to some extent by the formation of inhomogeneous Fermi surface structure.  A classification of filling-control Mott transition is possible with the criterion of $T_{ch}>0$ or formally $T_{ch}\le 0$~\cite{DHLee}.
    
%ƒ'ƒbƒg"]ˆÚ'Æ"ñŒ³‡‹àAƒCƒWƒ"ƒO"]ˆÚ'̂'Ȃª'è
\subsection{Mott Transition and Gas-Liquid Transition, Origin of Attraction}
If the transition between the Mott insulator and metals occur through the mechanism similar to the gas-liquid transition or the binary alloy transition, the above phenomenological and classical GL free energy should capture the essence.  In the gas-liquid-type transition, the presence of the attractive interaction of particles is known to be crucial.  In fact, the liquid phase does not exist without the attractive interaction. 
When the particle-particle interaction is attractive, we have a negative coefficient for the $n^2$ term of the free-energy expansion with respect to the density $n$ in Eq.(\ref{1}).  This results in the negative curvature in the free energy as a function of $n$ and this necessarily causes the phase separation (or two-phase coexistence) and the first-order transition, because the negative curvature (convex curve) makes it possible to draw a common tangent between two phase-separated densities.

In case of the transition to the band insulator, the carrier interaction is primarily repulsive and the picture of the gas-liquid transition does not apply straightforwardly.  On the other hand, the Mott insulator and charge order has a stability because of the commensurability effect of the electrons with underlying periodic lattice.  This commensurability extraordinarily lowers the energy of the insulating phase as compared to the metallic phase.  This lowering makes the kink structure for the energy at commensurate fillings as a function of the filling.  The question whether the attractive interaction exists or not as in the classical liquid phase is a highly nontrivial issue.  

%ˆø—Í'Ì‹NŒ¹Aƒ_ƒuƒƒ"ƒzƒƒ"ˆø—Í
% doubloin-holon'Ì‹­'¢Ã"dˆø—Í'ÆdoublonŠÔ'ÌŽã'¢'ŠŒÝì—pA'e'b'l's'Æ'a'b'l's'̈Ⴂ
% doublon-holon'©"›ó'ԁˆø—́AAA‰t'Š‹C'Š"]ˆÚ'Ƃ̃AƒiƒƒW[
However, in case of the bandwidth-control transition, the existence of the attractive interaction is rather obvious.  The Mott insulating phase is characterized by the doubly occupied site (doublon) and the empty site (holon) forming the bound state while the metallic phase is characterized by the unbinding of the doublon and the holon~\cite{RMP}.  The attractive interaction of a doublon and a holon clearly exists if $U$ is large, because the single doublon and holon approximately cost the energy $2U$ while the cost vanishes when they annihilate in pair.   The attractive interaction $E_b$ is $2U$ if the kinetic energy is ignored, and can be substantially less if the kinetic energy effect is considered. Then the term proportional to $n_dn_h$ has a negative coefficient in the free energy at $T=0$. If the filling is fixed at half filling, we have a constraint $n_d=n_h$.  After the transformation of the variables from $n_d$ and $n_h$ to $\nu$ and $\delta$, the negative curvature of the free energy as a function of $\nu$ appears under the constraint of $\delta=0$. This yields $A_dT_{cd}=E_b$.    If quantum effects can be ignored, the critical point thus obtained at $T=T_{cd}$ represents the classical Ising nature of the transition. 

The doublon-holon attractive interaction is enlarged by a stronger antiferromagnetic correlation because the doublon and holon disturb the antiferromagnetic background and the energy is lowered if they less disturb by the binding.  The order of the transition should be determined from the ratio of the holon-doublon interaction to the formation energy of the holon or doublon, which is basically scaled by $U$.  This is reminiscent of the conversion of the Berezinsky-Kosterlitz-Thouless tansition of the continuous type to the first-order transition when the vortex core energy is reduced relative to the vortex-antivortex attraction~\cite{Saito}.  Weaker first-order transition observed for larger geometrical frustration in a numerical study~\cite{Morita} of the Hubbard model on an anisotropic triangular lattice shown in Fig.~\ref{Fig1}B is explained from the weaker doublon-holon attraction because of the weaker antiferromagnetic background.

In case of the filling-control transition, it is much more subtle as we further discuss later.
The filling-control transition is realized by controlling the doping concentration, $\delta_c$ or $\delta$. The interaction between two holons is a nontrivial issue and should be quite different from the doublon-holon interaction. As we mentioned above, the stability of the Mott insulator is ascribed to the fact that the electron filling is one per lattice site, which is lost if the filling deviates from the Mott insulator.  The energy at half filling is extraordinarily lowered due to the commensurability.  If the states deviating from half filling remain uncorrelated metal, it necessarily causes a jump of the energy because of the loss of the energy due to the $U$ term, and it trivially leads to the phase separation if the filling deviates from commensurate filling.  This means the attractive holon-holon interaction.  In the underdoped region, the real metallic state is realized by compromising the correlation effect, and the energy in this region becomes much lower than that calculated from the free fermion state.  Then the jump in energy as a function of the filling is transformed to a continuous function with a kink and the negative curveture is much reduced or even may vanish. The origin of the attractive holon-holon interaction, however, if it exists, ultimately traces back to this commensurability force.    In any case, at least the holon-holon attractive interaction should be much weaker than the doublon-holon attraction, which explains the continuous character of the filling-control transition.  

The present observation suggests that the attractive interaction of holons or holons and doublons may be controlled to some extent. By this control, it may be possible to drive the system to the quantum criticality. At the quantum criticality, the critical end point of the first-order transition (or phase separation) occurs just at zero temperature.  

%"'lŒvŽZŒ‹‰Ê'Æ'Ì"äŠr
%"'lŒvŽZ'Å'Ì'e'b'l's'Æ'a'b'l's'̈ႢA
\section{Comparison with Numerical and Experimental Results}
\subsection{Numerical Results}

The insight from the GL expansion can be further analyzed with the help of recent numerical results of the Hubbard model.
In case of the bandwidth-control transition, the dynamical mean-field theory of the Hubbard model indicates that the transition is of the first order below a finite critical temperature.  The critical point is, moreover, characterized by the mean field exponent of the Ising transition~\cite{Kotliar,Kotliar2}.  Therefore, the GL mean-field picture seems to apply as the mean-field picture of the Mott transition.  

\begin{figure}
\begin{center}
%\epsfxsize=7cm
%\epsfbox{supi.eps}
\includegraphics[
width=2.5609in
]%
{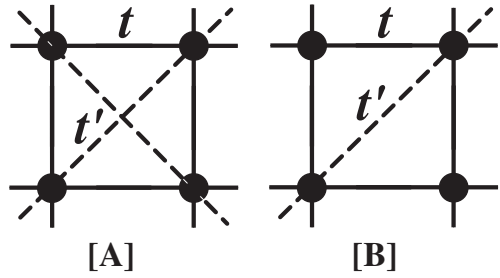}%
\caption{Lattice structure of geometrically frustrated lattices [A] on a square lattice and [B] on an anisotropic triangular lattice.  The nearest- and next-nearest-neighbor transfers are denoted by $t$ (solid bonds) and $t'$ (dashed bonds), respectively.}
\label{Fig1}
\end{center}
\end{figure}

In the two-dimensional Hubbard model, the path-integral renormalization group (PIRG) study~\cite{Kashima0,ImadaKashima} clearly shows that the first-order transition is retained in two dimensions with a jump of the double occupancy $n_d$ at $T=0$~\cite{Kashima1,Morita}.  Such bandwidth-control transitions in two dimensions are realized by introducing next-nearest neighbor transfers in addition to the nearest-neighbor transfer in the Hubbard model as is illustrated in Fig.~\ref{Fig1}.
The correlator projection method~\cite{Onoda}, in which the momentum dependence of the self-energy is taken into account in the dynamical mean-field theory, also consistently shows the first-order transition with a jump of the double occupancy and the finite temperature critical point in the same model~\cite{Onoda}.
As is mentioned above, the first-order transition for the bandwidth control is a natural consequence of strong doublon-holon attraction.  The first-order jump decreases with the decrease in the antiferromagnetic correlation by introducing frustration effects.  This is also consistent with the picture in the previous section.  

In case of the filling-control transition, the numerical results suggest that the order of the transition is more subtle in accordance with the analysis in the previous section.  The dynamical mean-field theory appears to show a presumable phase separation (only in the region of very small doping less than 1\%) at finite temperature while the single-band model calculation shows the continuous transition at $T=0$.  A small overestimate of the holon-holon interaction by some approximation easily alters the transition from continuous to the first order because an infinitesimally small overestimate transforms the diverging charge susceptibility at $\delta=\delta_s=0$ to that at a nonzero $\delta$ and leads to the phase separation. This issue will be discussed later.
It is well known that the mean-field theory may have a tendency to give a fictitious first-order transition because of the ignorance of the fluctuation effects near the transition.  
Small cluster studies on two-dimensional lattice appeared to show a phase separation~\cite{Emery}, but it turned out to be an artifact and result from overlooking the shell effect seen in the finite size clusters~\cite{FurukawaImada,FurukawaImada2}. 
In the quantum Monte Carlo~\cite{FurukawaImada,FurukawaImada2,Kohno} and PIRG~\cite{Watanabe} studies on the two-dimensional lattices, the charge susceptibility divergence appears to occur at zero temperature at the transition point $\delta=\delta_c=0$. The divergence occurs as $\chi_c \propto \delta^{-1}$.   
The diverging charge susceptibility at $T=0$ in fact means that $T_{ch}=0$ is just satisfied in Eq.(\ref{GL3}). 
This does not exclude very small but nonzero $T_{ch}$ and a small region of the phase separation if it is beyond the numerical accuracy.
In addition, the diverging charge susceptibility at $\delta=0$ suggests that the  critical temperature could be easily driven to a small but positive and nonzero value depending on the models, approximations and dimensionality.
At least within the Hubbard model, however, the filling-control transition in finite dimensional systems shows no evidence for the phase separation within the numerical accuracy. Instead the filling-control transition in two dimensions shows continuous character with diverging compressibility at vanishing doping concentration $\delta$. 
The numerical results indicate a general agreement that the bandwidth-control transition is of the first order while the filling-control transition shows a continuous nature with $T_c$ being nearly vanishing.

%VŽšŒ^'Š‹«ŠE
Recently, it has been clarified that this contrasted behavior between bandwidth-control and filling-control transitions is consistent with the peculiar V-shape structure of the phase boundary between the Mott insulator and metal in the plane of $U/t$ and the chemical potential~\cite{Watanabe}.

%ŽÀŒ±Œ‹‰Ê
\subsection{Comparison with Experimental Results}
%ŽÀŒ±'Ƃ̈ê'vA'g'…'RA'd's‰–A'u'Q'n'RAcuprate'⑼'Ì's'l'n'Ì–â'è
In strongly correlated electron systems as in transition metal oxides and organic compounds, particularly near their Mott insulating phases, various competing orders and their fluctuations are universally observed and the subject of intensive studies in the recent two decades~\cite{RMP}.  We will clarify later that such severe competitions at least partially originate from the tendency toward the underlying criticality of the Mott transition mentioned above.   To understand the striking complexity of these competing orders and their fluctuations as compared to weakly correlated systems, it is  important to understand the underlying mechanism and physics of the criticality. 

Very recently the first-order Mott transition with a critical point at nonzero $T$ in V$_2$O$_3$ has been reanalyzed in detail~\cite{Limlette}.  This bandwidth-control transition is consistent with the above GL expansion and numerical results.

In addition to enormous number of studies in transition metal compounds, we note two interesting recent experimental studies, one on the organic compounds and the other on the adsorbed monolayer $^3$He on graphite~\cite{Saunders,Fukuyama}.  Fournier {\it et al.}~\cite{Fournier} and Kagawa et al.~\cite{Kanoda} have shown that a first-order transition between the Mott insulator and a metal (at low temperatures, a superconductor) takes place in an organic material, $\kappa-$ET compound, and its first-order boundary starting from zero temperature extends up to the critical end point at around 34K in the plane of pressure and temperature. In this study, the pressure controls the relative bandwidth to the interaction strength while the electron filling is fixed, and therefore, a bandwidth-control transition is realized. In fact, according to the extended H\"uckel calculation
for $\kappa$-type ET compound~\cite{Mori}, a minimal model may be the single-band Hubbard model at half filling on an anisotropic triangular lattice defined in Eq.(\ref{3.1}) on the lattice structure given in Fig.~\ref{Fig1}[B].  As we already discussed in the previous subsection, this first-order transition with the critical end point was predicted~\cite{Morita,Onoda} for this lattice structure of the Hubbard model before these experiments.  

On the other hand, the $^3$He monolayer adsorbed on graphite substrate shows a typical filling-control transition to the Mott insulating state by changing the  $^3$He density. The Mott insulating state is realized at the commensurate density with the periodic potential formed by the underlying layer.  When the density approaches the Mott insulating state in the liquid phase, it shows striking increase and critical divergence of the effective mass probed by the specific heat coefficient and the magnetic susceptibility~\cite{Saunders,Fukuyama}.  The liquid phase nicely obeys the Fermi liquid criterion. Although a possible weak first-order character (namely phase coexistence or phase separation) cannot be excluded, this critical divergence implies the basic continuous character of the filling-control transition in contrast with the bandwidth-control transition observed in the organic compounds.    

Of course, it is questionable that the $^3$He system can be described by the Hubbard model and more realistic model with hard-core repulsion should be examined for a quantitative analysis.  It is also conceivable that the order of the transition may depend on the relevant model as we mentioned above.  However, it is also important to realize that the continuous-type transition for the filling control can occur with a critical divergence of fluctuations.   

These two different experiments performed on relatively clean and simple systems as compared to transition metal oxides show a similar contrast between bandwidth-control and filling-control transitions to numerical studies which we mentioned above. 

%—ÊŽqŒø‰Ê'̏d—v«    
\section{Quantum Effect}  
%â'Ηë"x'Å'Ì‹à'®â‰'Ì"]ˆÚA"d‰×Š´Žó—¦'ƃhƒ‹[ƒfd'Ý
\subsection{Mott Transition at T=0}
Insulators are characterized by vanishing conductivity at zero temperature.  Among various types of insulators, band insulators and Mott insulators are both distinguished from metals by vanishing Drude weight and vanishing charge susceptibility (compressibility) at zero temperature~\cite{RMP}.  As compared with phase transitions with symmetry breakings such as magnetic orders, the transitions from metals to the band insulators and the Mott insulators have been considered to be associated with no symmetry breaking by themselves.  However, the Drude weight and the charge susceptibility are regarded as the stiffnesses to twisted boundary conditions of the wavefunction phases~\cite{Kohn, Fisher, RMP}. The Drude weight is the stiffness to the twist of the phase of the wavefunction in the spatial direction, while the charge susceptibility is the stiffness to the temporal direction in the path-integral formalism.   In this sense, metals in a perfect crystal may be interpreted by the symmetry breaking of the phase of the spatially extended electron's wavefunction as compared to the localized and incoherent wavefunction in the insulator.  At zero temperature, insulators cannot be adiabatically continued to a metal and the distinction of metals and insulators is well defined. Then two phases have to be clearly separated by a phase boundary. On the other hand, such distinction is not clear at finite temperatures, because, strictly speaking, the wavefunction coherence is immediately destroyed away from $T=0$. 

\subsection{Coexistence with Fermi Degeneracy}  
The Mott transition discussed so far from the GL mean-field picture is essentially described as a classical one by ignoring quantum effects. The first-order metal-insulator transition can occur even without quantum degeneracy.  
If the Fermi degeneracy temperature near the first-order transition is much lower than the critical temperature $T_{ch}$ or $T_{cd}$, the quantum effects can be ignored.  In the opposite case, however, the quantum effects must be considered beyond the above framework.
The quantum effects have to be seriously considered when the Fermi degeneracy temperature becomes comparable or higher than the critical temperature of the Mott transition. 
Even when the bare Fermi temperature is high, the effective Fermi temperature is suppressed near the Mott critical point because approaching the continuous transition to the insulator should suppress the emergence of the Fermi degeneracy. However, the Fermi degeneracy may still coexist with the critical fluctuation in the metallic side near $T_c$.  This should be particularly eminent when $T_c$ is suppressed to zero.
  
In the Ising-transition picture~\cite{Goldenfeld}, the charge susceptibility $\chi_c$ diverges at the critical temperature as $\chi_c \propto (T-T_c)^{-\gamma}$ and $\chi_c \propto \delta^{-\zeta}$ with the exponents $\gamma$ and $\zeta$ given above.
The transition is characterized by these simple exponents with the hyperscaling assumption being satisfied below the upper critical dimension $d_u=4$.
The system has a single length scale $\xi$ which diverges at $T=T_c$.  In the present context, $\xi$ expresses the density correlation length or doublon density correlation length.  

% Œo˜HÏ•ª'É'æ'élŽ@A"®"I—ÕŠEŽw"'̏d—v«A
In the quantum region, however, we have to consider quantum dynamics.  This can be done by considering the path-integral formalism, where the imaginary time direction must be additionally considered in addition to the real spatial dimension. The time scale $\omega^{-1}$ diverges as $\omega^{-1} \propto \xi^{-z}$ in addition to the divergence of the spatial correlation length $\xi$.   The quantum dynamics is characterized by the dynamical exponent $z$.

% DMFT'Æ'ÌŠÖŒW@	'fChybridization function'Æ'ÌŠÖŒWiTc^d'ÆTc^n'Í"¯'¶'©Hj
In the mean-field level, the local quantum effect can be taken into account by the dynamical mean-field theory. The critical point of the metal-insulator transition is signalled by $\del^2 F/\del {\Delta}^2 =0$, where the hybridization function ${\Delta}$ is the conjugate variable to the single-particle local Green function $G$ and defined from ${\Delta}=\del F/\del G$.

\begin{figure}
\begin{center}
%\epsfxsize=7cm
%\epsfbox{supi.eps}
\includegraphics[
width=3.5609in
]%
{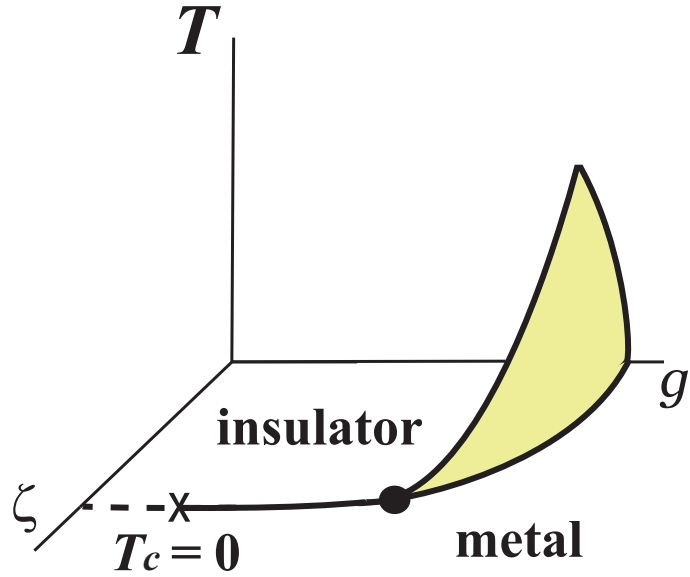}%
\caption{Schematic phase diagram of metals and insulators in the plane of a control parameter $\zeta$ such as $U$ or $\mu$, and $g$, which additionally controls the interactions of holons and doublons.  The continuous transition curve $T_c$ meets zero temperature at the solid circle and then $T_c$ stays at zero with further decrease in $g$.  We call the solid circle point {\it marginal quantum critical point}. The shaded surface represents the first-order transition. Along the $Tc=0$ curve, the coefficient of the quadratic dispersion, $a$, vanishes in the solid part, while it remains positive in the dashed part.  Thus the compressibility diverges in the solid part of the $T_c$ curve.  We call the region between the cross point and the solid circle (the marginal quantum critical point), {\it marginal quantum critical region}.}
\label{Fig2}
\end{center}
\end{figure}

%'±'̐ß(—ÊŽqŒø‰Êj'ÌŒ¤‹†"®‹@
% "ñŽŸ‹È—¦'ª³'̂܂܁iŒ`Ž®"I'É's'ƒ'ª•‰j'Æ'¢'¤'±'Æ'Í' 'é'©H|||ƒ{ƒ\ƒ"'Ì'Æ'«'́HAƒtƒFƒ‹ƒ~ƒIƒ"'Ì'Æ'«'̓oƒ"ƒhâ‰'Ì"]ˆÚ'Å'¶Ý'µ'Ä'¢'éBŒÃ"TŒn'Å'Ì's'ƒ'O'É'Š"–'·'é•t‹ß'©'çË—Í'¤'́H
As mentioned above, however, the mean-field theory may overestimate the tendency to the first-order transition, which may mask the quantum effects because of a resultant high critical temperature.  The order of the transition and the amplitude of the critical temperature may depend on the details of models, and materials. 
When the critical temperature $T_c$ is high, as we mentioned, the Mott transition is essentially classical and the mapping to the Ising transition holds.  One can suppress $T_c$ by controlling a parameter through the interaction of holons (or holons and doublons).  This corresponds to control $T_{ch}$ or $T_{cd}$ to zero. One might think that the critical temperature could become formally negative when one controls the parameter further.  This literally means that in the GL expansion of Eqs.(\ref{GL1}) and (\ref{GL3}), the metal-insulator transition disappears in this region.  However, this cannot happen because metals and insulators have to be clearly separated at zero temperature and the phase boundary cannot terminate as we have mentioned above.  This necessarily leads to the breakdown of the classical GL expansion or in other words, the transition always exists and the critical temperature stays just at zero temperature through the further parameter control.  As we see in Fig.\ref{Fig2}, this generates a $T_c=0$ line, where the transition would be far from the gas-liquid picture and the transition stays as the continuous quantum phase transition.  In this sense, this quantum criticality is different from the conventional quantum critical point. In the conventional case with a symmetry breaking at finite temperatures, the continuous critical point at $T \ne 0$ terminates at the quantum critical point without first-order transitions and without a $T_c=0$ line.  In the present case, at $T=0$, the $T_c=0$ line terminates at {\it marginal quantum critical point}, where the surface of the first-order transition starts opening with the nonzero critical temperature for the larger $g$.  Hereafter, our focus is largely on the nature around this marginal quantum critical point.  

% —ÕŠE‰·"x'ªƒ[ƒ'̃vƒƒgƒ^ƒCƒv'̍lŽ@Aƒ—˜_"I'É–Ê"''¢'¾'¯'Å'È'­A'e'b'l's"'lŒvŽZ'àŽxŽ'·'é
% ˆø—Í'ªŽã'­'Ä—ÊŽq—̈æ'É"ü'Á'Ä'«'½'Æ'«A,,ª–{"I–â'è
Here, we discuss how the quantum effect alters the transition by assuming the region that the critical temperature is low or even zero.  This corresponds to the region of vanishing effective interaction for holon or doublon in the GL expansion. 
When the critical temperature becomes low, the GL expansion tells that the charge fluctuation becomes diverging accompanied by the quantum degeneracy, which becomes beyond the scope of the form (\ref{GL1}) and (\ref{GL3}) .  
The numerical results of the filling-control transition in two dimensions discussed above indeed suggest a continuous transition at zero temperature with the diverging charge susceptibility, which is consistent with $T_{ch}=0$ and vanishing effective interaction between holons.  Therefore, this is certainly a realistic possibility.

% ƒ~ƒNƒ'ȃnƒ~ƒ‹ƒgƒjƒAƒ"AŒJ'荞'܂ꂽ1—±Žq—ã‹NA'Ž"WŠJ'ÌŠÔ'ÌŠÖŒW
%ˆê—±Žq•ªŽU'ƃ‰ƒ"ƒ_ƒE"WŠJ'ÌŠÖŒW
\subsection{Phenomenological Construction}
Here we formulate a phenomenological theory of the Mott transition in the quantum region.
When the path integral formalism with the imaginary time $\tau$ is employed, the free energy is formally expressed by using the Hubbard model as 
% ƒnƒo[ƒh–ÍŒ^'̃AƒNƒVƒ‡ƒ"
\begin{eqnarray}
F &=& -{\rm ln} Z/\beta N \\
Z &=& \int \prod_i {\cal D}\phi_i(\tau){\cal D}\phi_i^*(\tau){\rm e}^{-S/\hbar}, \\
S &=& \int_0^{\beta \hbar}{\rm d}\tau \sum_i \phi_i^* \hbar \del_{\tau} \phi_i
+ \int_0^{\beta \hbar}{\rm d}\tau H(\phi^*,\phi), \\
\label{PART1}
\end{eqnarray}
where the Hubbard hamiltonian $H$ for $N$ site system is rewritten by using the Grassmann variables $\phi_i$ and $\phi_i^*$ at the site $i$.

In the metallic phase except in one dimension, the adiabatic continuity of the Fermi liquid allows the description of the free energy by the renormalized single particle at low energies, where the higher-order terms are renormalized to the single-particle coefficient. 
In the insulating side, the action may also be given from a quasiparticle description with a gap $\Delta_c$ as
\begin{equation}
S=\sum_{i,n} \phi^* (-i\omega_n + \Delta_c + E(q,k))\phi
\label{Disp1}
\end{equation}
with the Matsubara frequency $\omega_n$,
and $E(q,k)$ is assumed to satisfy $E(q,k) \ge 0$ and vanishes at the gap edge.  Around the gap edge, we take the expansion in terms of $k$ as 
\begin{equation}
E(q,k)=a(q)k^2 + b(q)k^4+.....,  
\label{Disp2}
\end{equation}
where $k$ is the momentum coordinate perpendicular to the locus of $E(q,k)=0$  and $q$ denotes that parallel to the $E(q,k)=0$ surface. 
The gap edge is not necessarily an isolated point in the momentum space, but may be a line or a surface. Just at the transition point, $\Delta_c$ vanishes.
 
The coefficients $a$ and $b$ are regarded as renormalized values after the renormalization group treatment of the higher order terms in the expansion with respect to $\phi$.
In the presence of the $a$ term, the dynamical exponent $z$ characterized by the dispersion is given by $z=2$ as in the usual transition to the band insulator~\cite{RMPX}. For example in case of the filling-control transition, at $T=0$, the singular part of the free energy is expanded in terms of the density as $F=\delta^{(d+z)/d}$ as we see later, while the form Eq.(\ref{GL3}) is justified at $T>0$.  However, the hyperscaling assumption is still satisfied~\cite{RMPX} at $T=0$.  The transition between the band insulator and metals may also be interpreted from Eq.(\ref{GL3}) that $T_{ch}$ is kept formally negative for spatial dimensions $d\ge 2$ at $T>0$, while the singular form beyond Eq.(\ref{GL3}) appears only at $T=0$ as $F=\delta^{(d+2)/d}$.  

In this formalism, the attractive holon-holon (or holon-doublon) interaction should be renormalized and result in the flat dispersion in Eq.(\ref{Disp2}) at least in a part of the Brillouin zone when the phase separation or the first-order transition starts, because the localization by the phase separation (or diverging doublon susceptibility) should give the vanishing dispersion.  In other words, the coexisting phase or phase separation region is characterized by the region of diverging charge susceptibility or doublon susceptibility, which have to generate a diverging density of states at the gap edge in spatial dimension $d\ge 2$. This necessarily leads to $a \rightarrow 0$. 
Toward the marginal quantum critical point, the real part of the lowest order term $a$ continuously vanishes. This alters the dynamical exponent from 2 to 4, because the quadratic term proportional to $a$ vanishes and quartic term proportional to $b$ still remains in general. 
We note that in the GL description, this marginal behavior with $z=4$ is realized at the critical point, $T_{ch}$ or $T_{cd}$.  Strictly speaking, however, this argument is justified when the Fermi degeneracy is well developed and the quasiparticle description in the form (\ref{Disp1}) is valid.  This is satisfied around the marginal quantum critical point. 

Although the quasi-particle picture does not hold, it has numerically been shown that the dynamical exponent indeed becomes $z=4$ at the marginal quantum critical point in the one-dimensional Hubbard model with next-nearest-neighbor transfers~\cite{Nakano2}

If the gap edge given by $E(q,k)=0$ is composed of a line or a surface in the momentum space, the value $a$ in reality depends on the momentum along this line or surface.  When the system is metallized, this line or surface is expected to become the Fermi surface which satisfies the contained Luttinger volume.  Generically, the parameter $a$ depends on the momentum.  Therefore, when the parameter is controlled to the critical point, it is unlikely that $a$ vanishes simultaneously at all the points along the $E(q,k)=0$ surface in the metallic side.  On the contrary, generically the point with the smallest amplitude of $a$ becomes zero first as a special point of the $E=0$ surface.  Then a quartic dispersion appears at this special point of the $E=0$ surface when the system becomes marginal, which results in $z=4$.  
Therefore, the Mott critical point is also characterized by the evolution of an electron differentiation if the large Fermi surface is involved in the metallic side.  The singular differentiation generates a quartic dispersion at particular points of the Fermi surface coexisting with dispersive generic part.

This large dynamical exponent $z=4$ is indeed supported in a number of independent numerical calculations for the filling-control transition of the Hubbard model in two dimensions at $T=0$~\cite{FurukawaImada,FurukawaImada2,Imada1995,Assaad96,Tsunetsugu,Assaad99,Nakano}, which suggests that $T_{ch}$ is zero or very small in this case.  

We now consider the filling-control transition in more detail.  In the quantum region, the expansion in the form (\ref{GL1}) and (\ref{GL3}) is not justified any more.  However, the singular part of the free energy may still in principle be expanded at zero temperature as a function of the doping concentration $\delta$.  When the quadratic part of the dispersion proportional to $a$ around a point $q_1$ in eq.(\ref{Disp2}) remains, the lowest order term of the total energy measured from the insulator in this quasiparticle picture is given at $T=0$ from 
\begin{equation}
F=C a\delta^{(d+2)/d}
\label{Disp3}
\end{equation}
with a constant $C$ and therefore the charge susceptibility shows the scaling $\chi_c \equiv (\del^2 F/\del \delta^2)^{-1}\propto \delta^{1-2/d}$, which is the same as the transition to the band insulator. 
This is because
\begin{equation}
F\propto\int_0^{k_F}k^{d-1}k^2{\rm d}k
\label{Disp3-2}
\end{equation}
and
\begin{equation}
\delta\propto \int_0^{k_F}k^{d-1}{\rm d}k,
\label{Disp3-3}
\end{equation}
where we have integrated around $q_1$.
If a large Fermi surface which satisfies the Luttinger theorem appears immediately upon doping, one has to take that Fermi surface as the locus $E(q,k)=0$ and one gets
\begin{equation}
F=C \delta^{3}
\label{Disp3-1}
\end{equation}
because
\begin{equation}
F\propto\int_0^{k_F}k^2{\rm d}k
\label{Disp3-2}
\end{equation}
and
\begin{equation}
\delta\propto\int_0^{k_F}{\rm d}k,
\label{Disp3-3}
\end{equation}
where the integrations are performed in the region around the locus $E=0$.
However, in the marginal region near the marginal critical point with vanishing $a$ term at particular points $q_0$, we have the lowest order term
\begin{equation}
F=C b\delta^{(d+z)/d},
\label{Disp4}
\end{equation}
with $z=4$, which yields 
\begin{equation}
\chi_c \propto \delta^{1-z/d}.
\label{Disp4-1}
\end{equation}
The free energy near the Fermi ground state is therefore expressed as
\begin{equation}
F=C( a\delta^{(d+2)/d}+ b \delta^{(d+4)/d}),
\label{Disp5}
\end{equation}

Even in the case of the bandwidth-control transition, when one can regard the closing of the gap by hole doping around a point $q_{0h}$ and simultaneous particle doping around $q_{0p}$ with the constraint of keeping the electron density $n=1$, the above relation Eq.(\ref{Disp5}) may be replaced with $\nu$ as 
\begin{equation}
F= a\nu^{(d+2)/d}+ b \nu^{(d+4)/d},
\label{Disp6}
\end{equation}
because the ``doublon" and ``holon" concentration is nothing but the above self-doping concentration of particles and holes.  
However, the validity of this picture is limited because the large Fermi surface which satisfies the Luttinger theorem does not appear in this way.  In fact, the bandwidth-control transition seems to occur as the first-order transition at $T=0$ and the critical behavior may not exist.

A crucial point is that, at the marginal quantum critical point, the diverging charge susceptibility (or doublon susceptibility) driven by $z=4$ in the Fermi-degenerate region causes a multifurcating criticality as we discuss later. This instability is much stronger in the lower dimensions than in three dimensions as we see from Eq.(\ref{Disp4-1}).  In two dimensions, $\chi_c \propto \delta^{-1}$, while $\chi_c \propto \delta^{-1/3}$ in three dimensions.    

%ƒtƒBƒŠƒ"ƒO§Œä'ňø—Í'ÌŒ´ˆö'́HƒRƒƒ"ƒVƒ…ƒ‰ƒrƒŠƒeƒBH"½‹­Ž¥«'ä'炬H

\section{Two-Fluid Model}
Near the critical point of the Mott transition, $a(q)\ge 0$ has momentum dependence along the locus $E(q,k)=0$ and the marginal critical region is characterized by the vanishing $a(q)$ at particular points of the locus.  If it happens, the Fermi surface roughly consists of the two parts in the metallic side. One is the part where $a(q)$ shows rather large value, while the other is particular points where $a(q)$ gets smaller and smaller around such particular points $q=q_0$ as $a(q)=a_0\varepsilon + (q-q_0)^2$, where $\varepsilon$ measures the distance from the critical point in the control parameter, say $g$.  The contribution around the points $q_1$ where $a_1=a(q_1)$ is large gives the ground state energy as 
\begin{equation}
F= a_1\zeta^{(d+2)/d},
\label{Disp7}
\end{equation}
where $\zeta$ is either $\delta$ or $\nu$ depending on the filling or bandwidth control.  
The contribution from the area around $q_0$ is 
\begin{equation}
F= a_0\varepsilon\zeta^{(d+2)/d}+ b \zeta^{(d+4)/d}.
\label{Disp8}
\end{equation}
When the filling is controlled from a heavily doped region toward the marginal quantum critical point, there exists a crossover from the normal behavior, where the free enrgy is insensitive to $\delta$, 
to the critical proximity region, where 
\begin{equation}
F= b\delta^{(d+4)/d}.
\label{Disp9}
\end{equation}
In the normal region, the specific heat has a normal $\gamma$ value, which is essentially $\delta$ insensitive.
The specific heat may have a broad peak around a normal effective Fermi energy. 
With the decreasing doping concentration,
this high-temperature weight is transferred to the low-temperature region, where the critical proximity from Eq.(\ref{Disp9}) yields a quite different $\gamma$ value.  In this low-temperature structure, the $T$-linear behavior becomes limited to lower and lower temperatures with decreasing doping concentration. When $a_0$ vanishes, $\gamma$ diverges as $\gamma \propto 1/(a_0\varepsilon)$ while this $T$-linear behavior holds only in the region proportional to $a_0\varepsilon \delta^{2/d}+b\delta^{4/d}$ because the effective Fermi energy is scaled in this way. This low-temperature structure may continuously merge to the bahavior of the insulating phase, where the specific heat is determined from the spin entropy only and may have a peak structure at a characteristic spin-exchange energy.  In fact, this spin entropy contribution persists in the doped region and the region of the linear-$T$ specific heat should be limited to the lower energy than the peak of the spin contribution, when the doping concentration is low enough. 

Upon approaching the Mott critical point, the dispersion around $q_0$ becomes flatter and flatter and the damping effect may also become large.  In addition, scattering of quasiparticles in this momentum region may be responsible for the spin correlation in the insulating phase.  This flattness is the origin of the charge susceptibility divergence in Eq.(\ref{Disp4-1}).
Although the doped carriers are largely accepted in this region upon doping because of the flattness, the quasiparticle structure may not be clear because of the damping.  On the contrary, other part of the Fermi surface with dispersive quasiparticles accepts lower concentration of carriers while it shows more coherent quasiparticle structure.  Because of this electron differentiation depending on the momentum on the Fermi surface, the DC transport contributed from the dispersive region and the spin dynamics mainly arising from the flat region may behave as if they are separately evolving.  This is reminiscent of the phenomenological argument of ``hot" and ``cold" spots~\cite{Ioffe} in the high-$T_c$ cuprates.  

When $T_c$ becomes positive, the phase separation starts at $T<T_c$.  This phase separation starts from this flat region of the Fermi surface, while it can coexist with the metallic carriers in the dispersive region, if the carriers are also doped in the dispersive region.  If carriers are not doped into the dispersive part, the phase separation must be frozen at an early stage of the spinodal decomposition or the nucleation, because the long-ranged Coulomb repulsion prohibits further evolution of the phase separation.  When the holes are doped also into the dispersive part in addition to the flat part, the screening by the dispersive carriers may relax the constraint from the Coulomb repulsion.  The characteristic frozen domain size $\xi$ is $\xi \sim \sqrt{\epsilon \Delta E }/(\delta_0 e)$, where $\delta_0$ is the density difference between two phase separated area, $\Delta E$ is the free energy barrier between two minima, and $\epsilon$ is the effective dielectric constant.  The coexistence with the dispersive carrier increases $\epsilon$ through screening effects.  In a region of the patched real space, the carriers from the flat dispersion region are more dense and in the other region, coherent carriers from the dispersive region are more densely populated.  

The two-fluid model is a simplified picture for the coexistence of the dispersive and flat regions near the Mott critical point.  For more quantitative analyses, one has to consider the multi-fluid model that contains more than two components because the flat and dispersive regions are continuously interpolated.  
 
\section{Multi-furcation Instability}
Near the marginal quantum critical point $T_{ch}=0$, or more precisely in the region of $a=0$, the charge susceptibility diverges at $T=0$, while its coexistence with the Fermi degeneracy generates a drastic effect.  If these two coexist, the instabilities to various orders including magnetic, charge and superconducting phases are undoubtedly largely enhanced as compared to the existence of only one.  The antiferromagnetism, ferromagnetism, charge order and the superconducting order severely compete, because all of them easily diverge because of enhanced density of states arising from the proximity to the diverging charge susceptibility. 
The susceptibility at $T=0$ for the order parameter $u$ is given by 
\begin{equation}
\chi_u=\int {\rm d}E \rho(E) \frac{|\langle 0 | u^{\dagger}|\varphi_{E}\rangle|^2}{E-\omega},
\label{Multi1}
\end{equation}
which diverges because of the diverging density of states.  Here, $|\varphi_{E}\rangle$ is the eigenstate with the energy $E$, $|0\rangle$ is the ground state and $\rho(E)$ is the density of states.
The above orders are also competing with the tendency towards the phase separation. It turns out that such instability is enhanced in spatial dimensions lower than three because of Eq.(\ref{Disp4-1}), namely $\chi_c\propto \delta^{1-4/d}$.   Because the symmetry breaking is severely suppressed in one dimension, the systems with strong two-dimensional anisotropy with weak three-dimensional coupling may have the most drastic and largest energy scale of the severe competitions and instabilities.   

Near the Mott critical point, the instability towards various symmetry breakings occurs mainly at the particular points $p_0$ with the flattened dispersion.  
The amplitude of the instability for different orders sensitively depends on the details of the lattice structure, Fermi surface shape and band structure, because the position of $p_0$ may depend on details of systems.  We do not discuss individual differences of instabilities depending on the detailed distinction of systems.  An important point is that such multi-furcating instability generically starts developing and very sensitive dependence of fluctuation enhancement appears when the system approaches the marginal quantum critical point of the Mott transition. At least the diversity and complexity of the phenomena in experiments in this region~\cite{RMP} are understood from the underlying combination of the {\it Fermi degeneracy} and the {\it compressibility divergence} caused by a unique quantum criticality of the Mott transition. 

It is certainly possible that, under critical charge fluctuations of the Mott transition, some symmetry breaking occurs at finite temperatures before the real Mott critical point is reached.  For example, the antiferromagnetic order or superconducting order may appear before the real Mott critical point.  In this case, the transition to the Mott insulator occurs from such symmetry broken states and then the transition is ultimately rather analogous to the transition to the band insulator at $T=0$~\cite{Imada1993,Imada1994}.  For example, the transition between an antiferromagnetic insulator and an antiferromagnetic metal show similar behavior with the transition to the band insulator, because the folded Brillouin zone generated by the antiferromagnetic order yields the vanishing Fermi pocket at the metal-insulator transition point.   Such transitions between the symmetry broken phases are characterized by the vanishing number of carriers with a noncritical effective effective mass.  Instead, if the symmetry breaking does not occur, the real Mott criticality can be observed at $T=T_c$ and the diverging charge fluctuations cannnot be described by a mapping to a transition to the band insulator. Thus a classification scheme of two different types of transitions to the Mott insulators is useful~\cite{Imada1993,Imada1994}. One is the case with the Mott critical point observable and in the other, the critical point is masked by the symmetry breaking transition induced by the Mott criticality.

In this context of multi-furcation, instabilities of the metals near the Mott insulator have been examined from various standpoints, although the fundamental and unique structure of the Mott criticality with the divergent $\kappa$ has not necessarily been fully considered.  In fact, the controversies~\cite{RMP} in the literature all indicate that the stable phase near the Mott insulator very sensitively depends on models and approximations.  The divergent $\kappa$ as $\delta \rightarrow 0$ easily induces a real phase separation if a small hole-hole attraction is induced either by a real additional perturbation or by an artifact of the approximation.  
This means that $T_{ch}$ may be controlled to a substantially positive value if some additional interaction can be generated.  Then the phase separation can occur even at the classical level without the Fermi degeneracy.  This may happen, for example,  when the antiferromagnetic order is overestimated than the real one because it introduces an additional attraction of holes by hand.  A similar situation may appear when one takes the $t$-$J$ model with a rather large $J$.  The interaction of the holons should be determined in a self-consistent fashion with very subtle balance in the realistic electronic structure near half filling while the $J$ term controlled by hand very easily destroys this requirement.  

A more intriguing possibility among the candidates of the furcation is the superconductivity, where the coexistence of the quantum degeneracy is crucially important. There exist a large number of works on the possibility of superconducting phase without paying full attention on the unique character of the Mott transition discussed in this paper. Among them we briefly discuss some of the studies in relation to the present scope.   The enhancement of the $d$-wave superconducting correlations are signaled by tuning the level structure near the Fermi surface in numerical studies~\cite{Kuroki1}, which also suggests that the pairing is sensitively enhanced by emphasizing the degeneracy near the Fermi level.  The FLEX studies~\cite{Kuroki2} show an enhanced pairing induced by the effective attractive interaction coming from enhanced and overestimated antiferromagnetic fluctuations. 
The stabilization of the superconducting phase is also shown by taking a relatively large $J$ in the $t$-$J$ model in  variational or mean-field studies~\cite{AndersonRVB}. The $J$ term in the $t$-$J$ model explicitly enhances an attraction of two singlet electrons by hand, while an overestimate of the role of the antiferromagnetic correlations may also overestimate the stability of superconductivity.  

The stabilization of the superconducting state is also seen by taking a small additional perturbation of correlated hopping in numerically well controlled calculations~\cite{AssaadScalapino,TsunetsuguD}.  The correlated hopping term is in fact derived in the strong-coupling expansion of multi-band systems and can be significant because of the enhanced compressibility near the Mott transition~\cite{ImadaOnodaProc, Tsunetsugu}.  This superconductivity is driven by the kinetic energy gain, which is entirely different from the BCS mechanism~\cite{ImadaOnodaProc}. The kinetic energy gain arises because the quantum Mott criticality suppresses the kinetic energy of single particle through $z=4$. The attractive interaction of two pairs, each propagating coherently, is much reduced as compared to the attractive interaction of two holes, each coming from $q_0$ and $-q_0$.     
In other words, the marginal quantum critical region for paired electrons shifts from that of single particle so that the $T_c=0$ line extends for paired electrons than for unpaired electrons.  Because of this shift, the dispersion of pairs is not severely suppressed as compared with the single-particle dispersion.  An extreme sensitivity toward the superconducting instability may in reality be again a consequence of the diverging density of states near the Mott insulator.   In fact, when the effective interaction of the {\it mobile} holes becomes attractive, it is very natural to have a stabilization of the superconducting phase before the real phase separation (or spatial inhomogeneity) takes place when the Fermi degeneracy coexists. 
The kinetic pairing mechanism is expressed by the following effective Hamiltonian:
\begin{equation}
{\cal H}= \sum_{q,\sigma}E_1(q)c_{q\sigma}^{\dagger}c_{q\sigma}+\sum_q E_2(q)\Delta(q)^{\dagger}\Delta(q)
\label{KineticPair}
\end{equation}
with the pairing order parameter $\Delta(q)=f(q)c_{q\uparrow} c_{-q\downarrow}$.
In (\ref{KineticPair}), $E_1(q)$ has vanishing dispersion around $q_0$ only with the quartic term as $E_1(q)=(q-q_0)^4$, while $E_2(q)$ has a normal dispersion with the quadratic term. This drives the pairing of this form.  The form factor $f$ should be chosen so that the dispersion $E_2(q)$ becomes the largest. Because of this two-particle dispersion, dispersive pairing states are formed within the single-particle Mott gap. 

It should also be noted that if $a(q)$ becomes vanishing at four points, say, $q_0, -q_0, q_1$ and $-q_1$ on the Fermi surface in the Brillouin zone, it additionally favors the superconducting instability by the pairing of 
\begin{equation}
P_{q_0,q_1}=\langle c_{q_0\uparrow} c_{-q_0\downarrow}-c_{q_1\uparrow} c_{-q_1\downarrow}\rangle, 
\label{Pair}
\end{equation}
where the exchange interaction may work as attractive because of the negative phase factor in (\ref{Pair}).   This situation indeed occurs in the $d$-wave pairing realized in the high-$T_c$ cuprates at around $(\pi,0)$ and $(0,\pi)$.

The analysis in the present paper adds a new insight into the controversies for the determination of the ground state.  Most of the examined instabilities, 
including magnetic, charge, and pairing ones, are singularly enhanced by the coexistence of the Fermi degeneracy and the Mott criticality.  
All these instabilities induced by a small perturbation or crudeness of the approximation together with their controversies depending on the models and approximations certainly indicate the significance of underlying ``multi-furcating criticality".   The multi-unstable nature is also tightly connected with unusually suppressed coherence of quasiparticles in the flat-dispersion region around $q_0$, which is also a consequence of a large dynamical exponent $z=4$ derived in the underlying Mott transition.  The system becomes {\it almost} unstable to various directions in this case of the quantum criticality. 

Then we realize that we have to be very careful in concluding which type of the instabilities eventually wins since calculations suggest a strong degeneracy of various different phases under quantum fluctuations.  As we mentioned, if the critical point is suppressed to lower and lower temperatures as in the filling-control case, the instability is more enhanced because of the interplay with the Fermi degeneracy. Near the critical points of the Mott transition, whichever at $T_c=0$ and $T_c\ne 0$, the electronic state may have strong instabilities and they even further generate new subsequent instability mediated by enhanced fluctuations.  Without clarifying this subtle multi-furcating instabilities and competitions under the proximity to the compressibility divergence, the final result at zero temperature would not be reliable.  

We again note that the diverging charge susceptibility at the Mott critical point is connected with particular points of the Fermi surface, where the dispersion becomes flat.  This causes the ``differentiation" of electrons depending on the position of the Fermi surface.  It should be noted that this differentiation has a positive feedback effect.  The region of the Fermi surface, which has initially relatively stronger correlation effects, due either to the initial Fermi surface anisotropy or to anisotropic growth of fluctuations, allows stronger and stronger correlation effects.  This positive feedback appears because the dressed and slower quasipaticles with stronger damping capture the interaction effects more sensitively and strongly. Such positive feedback effects have numerically been seen in the flat dispersion around $(\pi,0)$ and $(0,\pi)$ points near the Fermi surface~\cite{Assaad99} of the Hubbard model and also discussed by the numerical renormalization group~\cite{furukawarice}.  The ``differentiation" in the momentum space may have a further possibility of the higher order structure at lower energy level, which results in a typical complex phenomena with a residual entropy retained.  
       
Moreover, it opens a new field of critical phenomena, where the originally single unstable fixed point generates ``daughters", namely, a multi-furcation of the subsequent and different unstable fixed points and the competitions of these fluctuations generate further lower-energy structure through a transient inhomogeneity in real and momentum space.  This is beyond the conventional scheme of the critical phenomena.  In fact, the analogous GL bifurcation point in the classical system generates simply a diverging susceptibility at the critical temperature, for example, in the gas-liquid and binary alloy mixture transitions.  At the classical level, this at most generates a nucleation and a spinodal decomposition regimes below the critical temperature and the physics is rather simple.  However, in the case of the Mott transition, the apparently similar problem turns out to be rich due to the interplay with the Fermi degeneracy, whose consequence may be regarded as a quantum emergence from the interplay of these two elements.  
Under a careful parameter control, an emergence of a complicated hierarchy is expected with lowering energy after severe competitions of various fluctuations before the ground state is reached.  
%The clarification of this complexity is one of the central issues in the condensed matter from the viewpoint of statistical physics.    

An attempt for searching a more enhanced instability towards such as superconductivity is also the subject of recent intensive studies, because the instability may depend on details of the models and a number of degrees of freedom~\cite{Aoki,KohnoImada}. 
Near the Mott critical point, emergence of the flat dispersion around $q_0$ points is a key element when one wishes to design an enhancement of desired orders, since the emergence of $q_0$ points may be properly controlled by system parameters.   By considering the above viewpoint of new hierarchy generation, this is a significant issue to be studied further.     

We have discussed only for the region near the Mott insulator at half filling of the Hubbard model.  However, in principle, a similar structure appears near any simple commensurate fillings of particle density.  It is known in many correlated electron systems that the Mott insulating state called the charge order is universally observed at a simple commensurate fillings such as $n=1/3$ and $1/4$, while it quantum mechanically melts away from such simple fractional densities~\cite{RMP}.  With the long-ranged Coulomb interaction, the melting is also driven by decreasing $m^*/\epsilon$ with $m^*$ and $\epsilon$ being the effective mass and the dielectric constant.  With increasing $m^*/\epsilon$, charge orderings are stabilized at more and more complicated fractional fillings and a structure of the devil's staircase appears as clarified recently by the PIRG method~\cite{Noda}.  In principle, a similar structure with divergent compressibility to the half-filled Mott insulator considered in this article may appear around each charge ordered states, which generates another complexity and hierarchy structure. The order of the charge order transition may again be determined from the ratio of the formation energy of the defect structure (charged defects) to the amplitude of defect-defect interaction.  An important difference from the quantum Mott transition is that the critical point is actually the marginal critical point in case of the charge order transition.  This is because that the charge-order transition necessarily accompanies the translational symmetry breaking. Therefore, the critical point as the termination point of the first-order transition must be the end point of the critical line. 

\section{Insight into High-Tc Cuprates and Other Strongly Correlated Materials}

It is useful to discuss the above consequences in relation to the observed experimental results.  
Various experimental results suggest that the Mott transitions in many of the transition metal compounds are rather close to the marginal quantum critical region especially for the filling-control transition. 
$^3$He adsorbed on a substrate also shows a critical behavior of the mass enhancement when the filling is controlled close to the Mott insulator (or registered phase).  

For the case of the bandwidth-control transition, the $\kappa$-ET type organic material~\cite{Fournier,Kanoda} and V$_2$O$_3$~\cite{Limlette} beautifully show the existence of the finite temperature critical point with the first-order transition below it.
 
For the filling-control transition, in particular high-$T_c$ cuprates show various indications which are consistent with the present picture.  
In the high-$Tc$ cuprates, flat dispersions are observed near $(\pi,0)$ and $(0,\pi)$ points.  
It is widely recognized that the flatness is beyond the conventional expectation from the van Hove singularity.   Such flattening is a natural consequence if the Mott critical point at $T=T_{ch}$ is located at low temperatures or at $T=0$ with an emergence of the point $q_0$ with vanishing quadratic dispersion, given by $a=0$ as we clarified in the previous sections.  
The actual $q_0$ positions responsible for the Mott criticality may depend on materials and models.  For example, Y and Bi compounds of the high-$T_c$ cuprates seem to be well modelled by a larger $t'$ than La based high-$T_c$ compounds when one takes the Hubbard model (\ref{3.1})~\cite{Andersen}.  Then the singular points $q_0$ may deviate from $(\pi,0)$ and $(0,\pi)$ points for larger $t'$.  Width of the critical region may be influenced by the amplitude of $t'$, while the existence of the criticality itself appears to be universal. In the electron doped cuprates, the critical behavior does not seem to exist because the doping within the antiferromagnetic ordered phase is equivalent to the doping into the band insulator.   

If the phase separation or spatial inhomogeneity of charge occurs, the critical point of the diverging compressibility shifts to a finite temperature as the critical termination point of the phase separation.  If this critical temperature is still in the region of the Fermi degeneracy in the metallic side, the same physics of the multi-furcating criticality we discuss in this article survives.  With the existence of the long-ranged Coulomb force, the phase separation is of course converted either to the charge order (or stripe) with a finite-length periodicity or to an inhomogeneity like the quenched and transient spinodal pattern, which has indeed been observed on the surface of the high-$T_c$ cuprates by the scanning tunnel microscope spectroscopy(STM and STS)~\cite{Pan}.  The observed spatial pattern of the inhomogeneity on the surface of the cuprate superconductors is very similar to the spinodal pattern widely observed in binary alloy systems~\cite{Spinodal,Cahn}.
The freezing of the spinodal pattern without the phase separation may be well explained by the long-ranged Coulomb repulsion.

The tendency for the phase separation is enhanced when itinerant electrons are coupled to classical degrees of freedom and thus the Mott critical temperature $T_c$ is enhanced. This happens in perovskite manganites, where the itinerant $e_g$ electrons are strongly coupled to localized $t_{2g}$ degrees of freedom through Hund's rule coupling~\cite{Mn1,Mn2}. It also couples to lattice degrees of freedom.    

The experimental results suggest that the stripe or inhomogeneity is actually rather easily stabilized if a corrugation potential from the lattice distortion or effects of random impurity potential, easily introduced by the doping, are present. This is again the consequence of the ``almost phase separated" nature realized by the criticality of the Mott transition with diverging $\kappa$ at the critical point.  If other symmetry breakings can be excluded, it is clear that the enhanced compressibility causes such inhomogeneous charge patterns.  
Although it is not clear whether the charge inhomogeneity is an intrinsic property of the high-Tc cuprates or it is stabilized just by the randomness and/or surface effects, the observed STM image strongly suggests that they are close to the spinodal or nucleation regions of the phase separation, because such inhomogeneous patterns have not been observed neither in heavily doped region of the cuprates nor in weakly correlated conventional metals even when the randomness is present.  Though the charge susceptibility $\chi_c(q,\omega)$ is suppressed strictly at $q=0$ and $\omega=0$, the experimental results suggests a substantial enhancement in the length scale smaller than the size of inhomogeneous pattern ($\sim 10$ nm). 

It should be noted that $^3$He system~\cite{Saunders,Fukuyama} does not have charge in contrast to the electronic systems.  He system seems to show continuous transition without a clear phase separation. The present analysis suggests that $T_{ch}$ is very low in this case.  It would be intriguing to analyze the Mott transition of $^3$He system on the substrate by the two-fluid or multi-fluid models discussed in Sec.6, as we can examine a wide temperature range.

The proximity of the Mott criticality around the marginal quantum critical point reveals, first, enhancement of the energy scale of the instabilities to various orders with strong competitions, and second, emergence of the electron differentiation in the momentum space together with a tendency for the inhomogeneity in the real space.  The electron diffferentiation generates flat bands around particular points of the Fermi surface while it strongly favors the anisotropic superconductivity.  All of these unusual properties are indeed observed in the high-$T_c$ cuprates.  The competing orders are a consequence of the mother criticality of the Mott transition, which generates many daughter instabilities.  Although various mechanisms of the superconductivity has been discussed in the literature, the criticality of the quantum Mott transition including the divergence of the charge susceptibility and charge fluctuations has not been fully studied. From the present insight, the primary mechanism of the cuprate superconductors is ascribed to the mother criticality of the quantum Mott transition, with the diverging charge fluctuations.  Secondary support of the pairing can be found in various fluctuations induced by the competitors such as the antiferromanetic fluctuations, while it does not drive the energy scale as high as in the Mott critical region. 

Further studies along this line would certainly be an intriguing issue.  More quantitative analysis of the multi-furcation due to the Mott criticality would be desired not only for searching superconductivity but also for other instabilities such as ferromagnetism and ferroelectricity.  
  
% •sƒ•¨Acorrugation potential'É'æ'é‰e‹¿
% 'Š•ª—£'Æ'·‹——£ƒN[ƒƒ"'É'æ'é—}§AƒXƒsƒm[ƒ_ƒ‹'Ì"€Œ‹A"d‰×'˜A•s‹Ïˆê«
% 
%$F=...$
% Šï"ŽŸ€'ªŒ»'ê'餁@h—LŒø"WŠJhAŒ»Û˜_'Æ'µ'ẴXƒP[ƒŠƒ"ƒO
% "d‰×Š´Žó—¦Aƒ_ƒuƒƒ"Š´Žó—¦,"d‰×'ä'炬'Ì"­ŽU''å,ó'Ô–§"x"­ŽU'ƃtƒFƒ‹ƒ~k'Þ'Ì'Šæ
% multi furcationA•ê'Æ–º@'½•ªŠòŒ^•ªŠòAŠeŽíŠ´Žó—¦'Ì‹­'¢"­ŽUA—\'ª•s‰Â"\«AƒoƒCƒAƒX'Ì–â'è
% —ÕŠE‰·"x'ª''·'¬'é'ƁAAAA
% self-organized criticality'Æ‹t
% ˆø—Í'Ì'‹ˆË'¶«A•'Ê'Í'‹ˆË'¶«'Ì'½'߁A•s‹Ïˆê«'Í•s‰Â"ð
% •Ö‹X"I'É"ñ''̗̈æ'É•ª'¯'čl'¦'éB
%
% ˆø—Í'Ì‹­'¢'Æ'±'ë'Í(1)'Š•ª—£AˆêŽŸ"]ˆÚ@(2)'´"`"±@(3)'x'ê'ătƒFƒ‹ƒ~—¬'Ì'Ö
%           @@@ƒtƒFƒ‹ƒ~–ÊŒ©'¦'È'¢A1—±ŽqƒŒƒxƒ‹'Í'©'Ȃ荂'¢'Æ'±'ë'É' 'é \> ARPES'Æ'ÌŠÖŒW
%                                                        ƒh[ƒsƒ"ƒOˆË'¶«
% ˆø—Í'ÌŽã'¢'Æ'±'ë'͍'‰·'©'çƒtƒFƒ‹ƒ~—¬'̉»
%   "ºŽ_‰»•¨'Æ'ÌŠÖŒW

%{\bf Discussions and Summary}
%%%%%%%%%%%%%%%%%%%%%%%%%%%%%%%%%%%%%%%%%%%%%%%%
%% BACKMATTER
%%%%%%%%%%%%%%%%%%%%%%%%%%%%%%%%%%%%%%%%%%%%%%%%

\noindent
{\bf Acknowledgments}

%\begin{acknowledgments}
The author would like to thank H. Nakano for useful discussions on unpublished results
prior to publication.

%\end{acknowledgments}

%% If the bibliography is produced without BibTeX comment out the

%\bibliography{sample}

\end{document}